\documentclass[fleqn,usenatbib]{mnras}
\usepackage[T1]{fontenc}

\DeclareRobustCommand{\VAN}[3]{#2}
\let\VANthebibliography\thebibliography
\def\thebibliography{\DeclareRobustCommand{\VAN}[3]{##3}\VANthebibliography}

\usepackage{graphicx}
\usepackage{dcolumn}
\usepackage{comment}
\usepackage{bm}
\usepackage{hyperref}
\usepackage{float}
\usepackage{amsmath}
\usepackage{amssymb}
\usepackage{mathtools}
\usepackage{newtxtext,newtxmath}

\newcommand{\md}{\mathrm{d}}

\newcommand{\bw}{\bm{w}}
\newcommand{\tsec}{t_\mathrm{sec}}
\newcommand{\bW}{\bm{W}}
\newcommand{\DF}{f}
\newcommand{\nn}{\nonumber}

\newcommand{\der}[2]{\frac{\md#1}{\md#2}}
\newcommand{\Msun}{M_\odot}
\newcommand{\alphaeff}{\alpha_\mathrm{eff}}

\usepackage{accents}
\newcommand*{\dt}[1]{\accentset{\mbox{\large\bfseries .}}{#1}}

\title[Eccentricity of wide binaries - I. Galactic tides]{Eccentricity dynamics of wide binaries - I. The effect of Galactic tides}

\author[S. Modak \& C. Hamilton]{
Shaunak Modak$^{1}$\thanks{E-mail: shaunakmodak@princeton.edu} and 
Chris Hamilton$^{2}$
\\
$^{1}$ Department of Astrophysical Sciences, Princeton University, 4 Ivy Lane, Princeton, NJ 08544, USA\\
$^{2}$ Institute for Advanced Study, Einstein Drive, Princeton, NJ 08540, USA\\
}

\date{Accepted XXX. Received YYY; in original form ZZZ}
\pubyear{2023}

\begin{document}
\graphicspath{{./}{figures/}}
\label{firstpage}
\pagerange{\pageref{firstpage}--\pageref{lastpage}}
\maketitle

\begin{abstract}
A major puzzle concerning the wide stellar binaries (semimajor axes $a\gtrsim 10^3$\,AU) in the Solar neighborhood is the origin of their observed superthermal eccentricity distribution function (DF), which is well-approximated by $P(e)\propto e^\alpha$ with $\alpha\approx 1.3$. This DF evolves under the combined influence of (i) tidal torques from the Galactic disk and (ii) scattering by passing stars, molecular clouds, and substructure. Recently, it was demonstrated that Galactic tides alone cannot produce a superthermal eccentricity DF from an initially isotropic, non-superthermal one, under the restrictive assumptions that the eccentricity DF was initially of power law form and then was rapidly phase-mixed toward a steady state by the tidal perturbation. In this paper we first prove analytically that this conclusion is valid at all times, regardless of these assumptions. We then adopt a thin Galactic disk model and numerically integrate the equations of motion for several ensembles of tidally perturbed wide binaries to study the time evolution in detail. We find that even non-power law DFs can be described by an effective power law index $\alpha_\mathrm{eff}$ which accurately characterizes both their initial and final states, and that a DF with initial (effective or exact) power law index $\alpha_\mathrm{i}$ is transformed by Galactic tides into another power law with index $\alpha_\mathrm{f}\approx (1+\alpha_\mathrm{i})/2$ on a timescale $ \sim 4\,\mathrm{Gyr}\,(a/10^4\mathrm{AU})^{-3/2}$. In a companion paper, we investigate separately the effect of stellar scattering. As the GAIA data continues to improve, these results will place strong constraints on wide binary formation channels.
\end{abstract}



\section{Introduction}
\label{sec:introduction}


Wide binaries are crucial tools for constraining the properties of dark matter substructure in the Solar neighborhood \citep{ramirez2022constraining} and in satellite galaxies \citep{penarrubia2016wide}, for ruling out the existence of MACHOs \citep{bahcall1985maximum, yoo2004end}, and for testing alternative theories of gravity \citep{hernandez2012wide, pittordis2019testing}. Such constraints are possible because wide binaries are simple dynamical systems which respond in a predictable way to gravitational forces, whether due to coherent Galactic tides or random scattering from stars, gas clouds, or substructure \citep{Weinberg1987, jiang2010evolution}.

Most dynamical studies of wide binaries in the literature have focused on the same metric: the binaries' semimajor axis (or rather, separation) distribution; or, even more crudely, the maximum separation distance of surviving binaries. In the past this approach was justified, since observations were unable to probe any other characteristic. However, in the age of GAIA, we can perform (statistical) measurements of the binaries' \textit{internal} phase space distribution functions (DFs), in particular their eccentricity DF, and combine this with measurements of their component masses, ages, Galactocentric kinematics, and more (e.g. \citealt{el2021million}). 

Crucially for this work, \cite{Tokovinin2020-go} and \cite{Hwang21} found that wide binaries (defined as those with projected separations $\geq 10^3$\,AU) have a superthermal eccentricity distribution, i.e. there is a strong enhancement at high $e$, and a corresponding deficit at low $e$, compared to the thermal distribution $P_\mathrm{th}(e) = 2e$.
In fact, \cite{Hwang21} found that the binaries with separations $\sim 10^3$\,AU were \textit{even more superthermal} than those with separations $\sim 10^4$\,AU. The origin of these superthermal DFs is not understood. Another unexplained observation is that ``twin'' wide binaries (those whose components have very similar masses) are almost all very eccentric, with $e \gtrsim 0.95$ \citep{hwang2022wide}. By combining various measurements of this kind, we should be able to place strong constraints on the properties of the Galactic environment in which wide binaries evolve, and/or on the formation mechanism(s) of the wide binaries themselves \citep{kouwenhoven2010formation, reipurth2012formation, lee2017formation, penarrubia2021creation, Rozner2023}.

The first step in such a program is to understand how the Galactic environment impacts binaries in a manner that goes beyond the classic studies of wide binary survival and semimajor axis distribution \citep{Weinberg1987, jiang2010evolution}. In this work, we will focus specifically on the origin of the observed superthermal eccentricity distribution for binaries with $a\gtrsim 10^3$\,AU. This program was initiated by \cite{Hamilton22} (hereafter H22), who studied the secular effect of Galactic disk tides alone (i.e. ignoring scattering) on the eccentricity and inclination distributions of (bound) wide binaries. The binary-disk system constitutes an effective three-body problem: Galactic tides drive secular oscillations in an individual binary's eccentricity and inclination, just like in the von Zeipel-Lidov-Kozai mechanism \citep{von1910application, Lidov1962, Kozai1962, Heisler1986-eb, Hamilton2019-jn, Hamilton2019-zl}, with oscillation timescale
\begin{equation}
    \label{eq:t_sec_numerical}
    t_\mathrm{sec} \sim 1\,\mathrm{Gyr}\,\left( \frac{\rho_0}{0.2 \Msun\,\mathrm{pc}^{-3}}\right)^{-1}
    \left(\frac{m_b}{\Msun}\right)^{1/2} \left(\frac{a}{10^4\,\mathrm{AU}}\right)^{-3/2},
\end{equation}
where $m_b$ is the binary's total mass and $\rho_0$ is the mass density in the Solar neighborhood. Upon applying this mechanism to an entire ensemble of initially isotropically-oriented binaries, H22 concluded that the observed superthermal DF cannot be produced by Galactic tides, unless the initial DF was \textit{even more} superthermal. However, H22's study was limited to initial eccentricity DFs of power law form, and, most questionably, to the ``phase-mixed limit'' in which Galactic tides have already driven the DF to a steady state. The phase-mixed assumption requires binaries to undergo multiple secular oscillations within the lifetime of the Galaxy, but since the timescale for these oscillations given by \eqref{eq:t_sec_numerical} is so long, this assumption may fail in practice.

In this work, we first relax several of the assumptions of H22, and show that in fact, a superthermal DF cannot be produced from an isotropic, non-superthermal one at \textit{any} time, nor for \textit{any} disk model, regardless of whether the intial DF is of power law form (\S\ref{sec:ensemble_evolution}). Rather, H22's conclusions are a generic and inevitable consequence of Hamiltonian dynamics (Liouville's theorem in particular). This means that the phase-mixed assumption is unnecessary. Next, we adopt the same disk model as in H22 (with more careful justification), and integrate the tidal equations of motion numerically (\S\ref{sec:Milky_Way}). This allows us to extend the results of H22 by tracking the eccentricity DFs over time and studying their convergence to a steady state. In \S\ref{sec:discussion}, we discuss the limitations of this study, and we summarize in \S\ref{sec:summary}.

In a companion paper (Hamilton \& Modak, in preparation; hereafter Paper II) we examine in detail the impact of stellar scattering upon wide binary DFs, and thereby place constraints on wide binary formation channels.


\section{Evolution of an ensemble of binaries in an external tidal field}
\label{sec:ensemble_evolution}


In this section we first recap the basic formalism for describing the dynamics of a single binary in a smooth, weak external tidal field which we will later take to be that of the Galactic disk (\S\ref{sec:dynamics}) --- for a detailed analysis of this problem, see \citet{Hamilton2019-jn, Hamilton2019-zl}. We then introduce the distribution functions that allow us to characterize the dynamical state of an ensemble of binaries, and the Vlasov equation which describes the ensemble's evolution (\S\ref{sec:vlasov}). Next, we define certain orbit-averaged DFs (\S\ref{sec:distribution_functions}), characterize what we mean by ``subthermal'' and ``superthermal'' eccentricity DFs (\S\ref{sec:subvssuper}), and discuss how any measured DF must be coarse-grained (\S\ref{sec:coarse_graining}). Finally, we bring all of these results together in order to place robust constraints on the possible evolution of eccentricity distributions (\S\ref{sec:proofs}).

We emphasize that unlike in H22, in this section we will not make any assumptions about the form of the external potential, truncate the tidal forces at any order, or restrict the discussion to a time-asymptotic phase-mixed state. Thus, most of the results derived here apply not only to wide stellar binaries in the Galaxy, but in fact to \textit{any} ensemble of Keplerian orbits evolving in an \textit{arbitrary} (smooth, weak, possibly time-dependent) tidal field. They are therefore equally valid for discussing e.g. the distribution of the trans-Neptunian objects perturbed by a hypothetical Planet Nine \citep{batygin2016evidence}, or the disk of young stars at the Galactic center \citep{haas2011secular, von2022young}. See the discussion in \S\ref{sec:discussion} for more details.


\subsection{Dynamics of a single binary}
\label{sec:dynamics}


Consider a binary with total mass $m_b$ whose Keplerian orbital motion (the ``inner'' orbit) is described by the semimajor axis $a$, eccentricity $e$, inclination $i$, longitude of ascending node $\Omega$, argument of pericenter $\omega$, and mean anomaly $\eta$. Here $i$ is measured relative to some fixed $(X, Y)$ plane, which we will later take to be the Galactic plane, and $\Omega$ is measured relative to the fixed $X$ axis. An alternative description of this inner orbit, which is more convenient for studying Hamiltonian dynamics, is provided by the Delaunay angles $\bm{\theta} = (\eta, \omega, \Omega)$ and their conjugate actions $\bm{I} = (L, J, J_z)$, with $L \equiv \sqrt{Gm_b a}$, $J \equiv L\sqrt{1-e^2}$, and $J_z \equiv J \cos i$. We will also find it convenient to introduce the quantities
\begin{align}
    \label{eq:def_j}
    j & \equiv J/L = \sqrt{1-e^2}, \\
    j_z & \equiv J_z/L = j\cos i
\end{align}
which correspond to the dimensionless total and $z$-component angular momenta of the binary respectively. With these definitions, $\omega \in [0, 2\pi)$, $\Omega \in [0, 2\pi)$, $j_z \in [-1, 1]$, and $j \in [|j_z|, 1]$.

Let the binary dynamics be determined by the Hamiltonian $H(\bm{\theta}, \bm{I}, t)$, which we leave arbitrary for now (a specific choice of $H$ encoding quadrupolar Galactic disk tides will be introduced in \S\ref{sec:Galactic_Disk}). Then the Delaunay variables evolve according to Hamilton's equations:
\begin{align}
    \label{eq:Hamiltons_Equations}
    \frac{\md \bm{\theta}}{\md t} = \frac{\partial H}{\partial \bm{I}},\  \frac{\md \bm{I}}{\md t} = -\frac{\partial H}{\partial \bm{\theta}.}
\end{align}
In the absence of external perturbations, $H = H_0 \equiv -Gm_b/2a$, and so $\eta$ evolves at the Keplerian mean motion rate $n \equiv \sqrt{Gm_b/a^3}$, while all other Delaunay variables remain constant.

Now let us impose an additional gravitational potential $\Phi$ which produces a tidal force on the binary. We assume $\Phi$ is sufficiently weak that it can alter the dynamics only on timescales much longer than the binary's inner orbital period $2\pi/n$, so we are justified in averaging the binary equations of motion over $\eta$. The resulting system of equations can be encapsulated in a Hamiltonian $H = H_0 + \delta H(\omega, \Omega, L, J, J_z, t)$. Since $\eta$ is not present in this Hamiltonian, $L$ (and therefore the binary's semimajor axis $a$) is conserved. The explicit time dependence of the Hamiltonian $\delta H$ allows for variations of the perturbation due to the binary's barycentric (``outer'') motion, e.g. as a wide binary oscillates vertically about the Galactic midplane. The potential $\Phi$ can in principle also have explicit time-dependence of its own (e.g. due to secular heating of the Galactic disk). The motion of a given binary through the $(\omega, \Omega, J, J_z)$ phase space may then be determined by plugging $H = H_0+\delta H$ into \eqref{eq:Hamiltons_Equations}.


\subsection{The Vlasov equation}
\label{sec:vlasov}


Consider a very large ensemble of binaries evolving under the same Hamiltonian $H$. We describe the ensemble using the phase-space DF $f(\bm{W}, t)$, where $\bm{W} \equiv (\bm{\theta}, \bm{I})$, such that the number of binaries with phase space coordinates in the volume $(\bm{W}, \bm{W}+\md \bm{W})$ at time $t$ is proportional to $f(\bm{W}, t)\,\md\bm{W}$. The dynamics of an individual binary is then given by Hamilton's equations \eqref{eq:Hamiltons_Equations}, and the resulting equation describing the evolution of the ensemble's DF $f$ is the Vlasov equation,
\begin{align}
    \label{eq:kinetic}
    \frac{\md \DF}{\md t} = \frac{\partial f}{\partial t} + \frac{\partial }{\partial \bm{W}} \cdot (\DF \dt{\bm{W}}) = 0,
\end{align}
where
\begin{align}
    \dt{\bm{W}} = \left(\frac{\partial H}{\partial \bm{I}}, -\frac{\partial H}{\partial \bm{\theta}} \right)
\end{align}
is the ``velocity'' of the phase space flow. Liouville's theorem guarantees that this flow is incompressible:
\begin{align}
    \frac{\partial}{\partial\bm{W}} \cdot \dt{\bm{W}} = 0.
    \label{eq:Liouville}
\end{align}

Equation \eqref{eq:kinetic} is all we need to evolve any initial DF $f(\bm{W},0)$ under the flow generated by $H$. However, to provide intuition for our upcoming results it is worth discussing a non-trivial quantity which is conserved under this evolution, namely
\begin{align}
    \label{eq:Casimir}
    C_2(t) \equiv \int \md \bW \, [f(\bW, t)]^2.
\end{align}
This functional of $f$ is sometimes called the quadratic Casimir, phasestrophy \citep{diamond2010modern}, or generalized entropy \citep{zhdankin2022generalized}, and is just one of an infinite family of Casimir invariants conserved by \eqref{eq:kinetic}\footnote{It is also an ``H-function'' in the sense of \cite{THL_1986}, but we refrain from using that terminology here.}. It is easy to show that $C_2$ is conserved:
\begin{align}
    \label{eq:C2_conserved}
    \der{C_2}{t} & = \int \md\bW\,  2f  \frac{\partial f}{\partial t} \nn = - \int \md\bW \, 2f \frac{\partial }{\partial \bW} \cdot (f\dt{\bW}) \\
    & = - \int \md\bW \frac{\partial }{\partial \bW} \cdot (f^2 \dt{\bW}) = 0,
\end{align}
where we have used Liouville's theorem \eqref{eq:Liouville} in the second and third equalities, and the final equality follows from the fact that $f^2\dt{\bW}$ must be periodic in angles. Note that this proof assumes nothing about the form of the Hamiltonian $H$, which could be arbitrarily complicated or time-dependent.

Heuristically, $C_2$ is a global measure of how non-uniform the DF is, and is minimized by a completely uniform DF $f = $ constant. Furthermore, if we split $f(\bW, t)$ into an ``averaged'' part and a ``fluctuation'' (though at this stage we do not have to define what we are averaging over, or require the fluctuation to be small)
\begin{align}
    f(\bW, t) = \overline{f}(\bW, t) + \delta f(\bW, t),
\end{align}
then $C_2$ has the appealing property that it splits cleanly into a part that depends on $\overline{f}$ and a part that depends on $\delta f$:
\begin{align}
    \label{eq:Casimir_Split_1}
    C_2(t) = \overline{C}_2(t) + \delta C_2(t),
\end{align}
where 
\begin{align}
    \label{eq:Casimir_Split_2}
    \overline{C}_2(t) & \equiv \int \md \bW [\overline{f}(\bW, t)]^2, \nonumber \\
    \delta C_2(t) & \equiv \int \md \bW [\delta f(\bW, t)]^2.
\end{align}
Thus, one can think of the ensemble's evolution in terms of a ``$C_2$ budget'' which can be transferred from the averaged part of the DF into the fluctuations and vice versa, but must be conserved in total. It is for this reason that we choose to study $C_2$ as opposed to other invariants such as the entropy, which do not have this ``splitting'' property.


\subsection{Orbit-averaged distribution functions}
\label{sec:distribution_functions}


Our assumption of weak tides (\S\ref{sec:dynamics}) means that each binary's action $L$ is conserved, so populations of binaries at each semimajor axis $a$ evolve separately. Of course, impulsive scattering can alter $a$ --- we discuss this possibility in \S\ref{sec:discussion}. At fixed $a$, the dynamical state of each binary is characterized by the masses of the constituents and the remaining (dimensionless) Delaunay variables $(\omega, \Omega, j, j_z)$. Moreover, the wide binaries of interest all have typical masses of $\sim 1\Msun$ \citep{Hwang21}, and small differences in mass should not have any significant effect on the dynamical evolution. Thus, for simplicity, throughout the rest of this work we will consider binaries that all have the same total mass $m_b$, so we can describe our ensemble of binaries using the reduced DF $f(\bw, t)$, where
\begin{align}
    \label{eq:def_DF}
    \bw \equiv (\omega, \Omega, j, j_z),
\end{align}
normalized such that $\int \md\bw f(\bw, t) = 1$. Thus, $f(\bw, t)\md \bw$ is the fraction of binaries in the ensemble with coordinates in the phase space volume $(\bw, \bw + \md\bw)$ at time $t$.

Using $f$, we can define the marginal distribution of dimensionless angular momenta
\begin{align}
    F(j, t) \equiv \int_{-j}^{j}\md j_z \int_0^{2\pi}\md\Omega \int_0^{2\pi}\md\omega f(\bw, t),
    \label{eq:marginal_DF}
\end{align}
which satisfies $\int_0^1 \md j F(j, t) = 1$, and the corresponding distribution of eccentricities,
\begin{align}
    P(e, t) \equiv \frac{e}{\sqrt{1-e^2}}F(\sqrt{1-e^2}, t),
    \label{eq:P_to_F}
\end{align}
which similarly satisfies $\int_0^1 \md e P(e, t) = 1$. Additionally, we define the marginal distribution of inclinations as 
\begin{align}
    \label{eq:marginal_cosi}
    N(\cos i, t) = \int_0^1 \md j \int_0^{2\pi} \md \Omega \int_0^{2\pi} \md \omega jf(\bw, t),
\end{align}
where we implicitly write $j_z = j\cos i$ in the argument of the DF in the integrand. This satisfies $\int_{-1}^{1}\md\cos i\, N = 1$.

In the special case that binary orientations are distributed isotropically, the DF will be uniform in $\omega$, $\Omega$, and $\cos i = j_z/j$. In this case $f$ may be considered a function of $j$ and $t$ only, so that
\begin{align}
    f_\mathrm{iso}(j, t) = \frac{F(j, t)}{8\pi^2j} = \frac{P(e, t)}{8\pi^2e}.
    \label{eq:f_isotropic}
\end{align}


\subsection{Classification of eccentricity distributions}
\label{sec:subvssuper}


We now discuss what precisely we mean by ``subthermal'' and ``superthermal'' DFs using the marginal eccentricity distribution $P(e)$ defined in equation \eqref{eq:P_to_F}.

One important class of eccentricity DFs is the power laws
\begin{align}
    \label{eq:ecc_power_law}
    P^{(\alpha)}(e) \equiv (1+\alpha)e^\alpha,
\end{align}
where $\alpha \geq 0$ is the power law index. These distributions are useful for their analytical simplicity, and often correspond well to observational data \citep{Hwang21}. The thermal distribution,
\begin{align}
    \label{eq:thermal}
    P_\mathrm{th}(e) = 2e,
\end{align}
is a special case of a power law eccentricity DF with $\alpha = 1$. A power law DF with $\alpha < 1$ is then naturally called ``subthermal,'' while one with $\alpha > 1$ is called ``superthermal.''

\begin{figure*}
    \centering
    \includegraphics[width=0.99\textwidth]{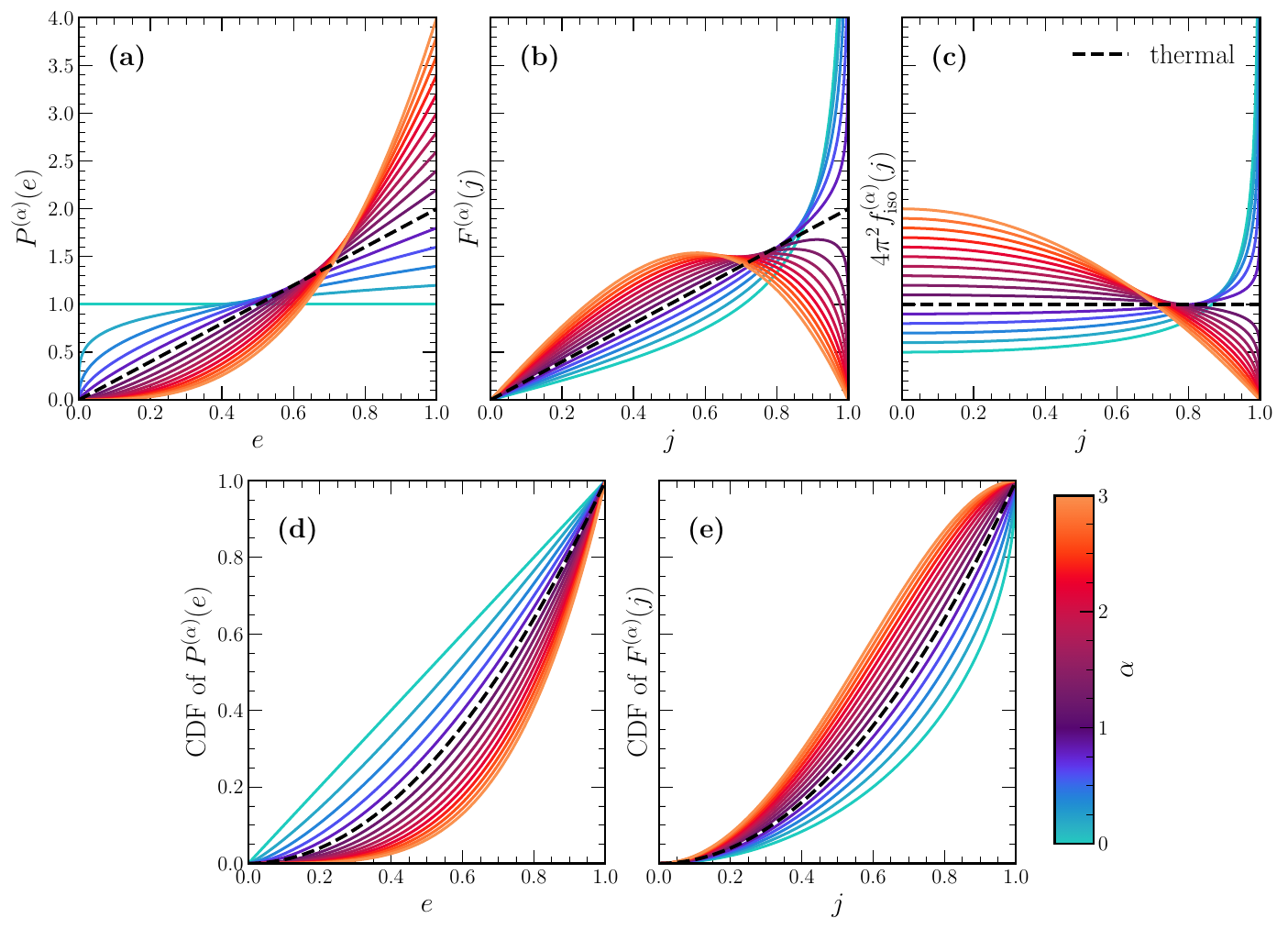}
    \caption{Panels (a) and (b) show the power law eccentricity distributions $P^{(\alpha)}(e)$ and the corresponding angular momentum DFs $F^{(\alpha)}(j)$ respectively, for sixteen evenly spaced values of power law index $\alpha$ from 0 to 3 inclusive (shown with different colored lines). Panel (c) shows the corresponding full DF $f^{(\alpha)}_\mathrm{iso}(j)$ given by equation \eqref{eq:f_isotropic}, assuming the binaries are oriented isotropically. Panels (d) and (e) show the CDFs of the distributions shown in panels (a) and (b) respectively.} In each panel the black dashed line indicates the thermal distribution ($\alpha = 1$).
    \label{fig:PFf_powerlaws}
\end{figure*}

In Figure \ref{fig:PFf_powerlaws}, we plot $P^{(\alpha)}(e)$ for various values of $\alpha$. We also plot the corresponding marginal DF $F^{(\alpha)}(j)$ following equation \eqref{eq:P_to_F}, as well as the full DF $f^{(\alpha)}_\mathrm{iso}(j)$ following equation \eqref{eq:f_isotropic}, which is valid if the binaries are distributed isotropically. For reference, in the lower panels we also plot the cumulative distribution functions (CDFs) that correspond to $P^{(\alpha)}(e)$ and $F^{(\alpha)}(j)$. We see that subthermal DFs have a surplus of binaries at low $e$ (high $j$) compared to the thermal distribution, and a deficit at high $e$ (low $j$). Superthermal DFs correspondingly show a surplus at high $e$ and a deficit at low $e$.

\begin{figure}
    \centering
    \includegraphics[width=0.48\textwidth]{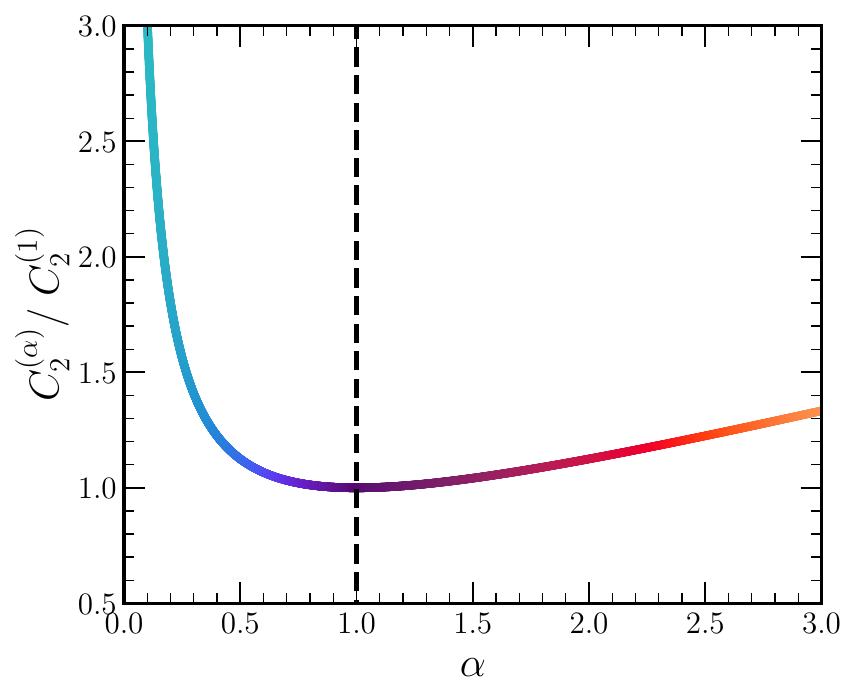}
    \caption{The functional $C_2$ given by equation \eqref{eq:Casimir} evaluated for isotropic power law DFs $f^{(\alpha)}_\mathrm{iso}$ as a function of the eccentricity DF power law index $\alpha$, normalized by the $C_2$ value for an isotropic thermal DF,  $C_2^{(1)} = 1/(4\pi^2)$. The color scale is as in Figure \ref{fig:PFf_powerlaws}. The black dashed line indicates the thermal power law index value $\alpha=1$.}
    \label{fig:C_alphas}
\end{figure}

In Figure \ref{fig:C_alphas} we plot the values of $C_2$ given by equation \eqref{eq:Casimir} for isotropic power law DFs $f^{(\alpha)}_\mathrm{iso}$ as a function of the eccentricity DF power law index $\alpha$, normalized by the value for the isotropic thermal DF, $C_2^{(1)} = (4\pi^2)^{-1}$. Since the isotropic thermal DF is uniform over the entire phase space, it naturally has a smaller $C_2$ value than any other possible distribution function.
Since $C_2$ is conserved it follows that an initially isotropic thermal DF will always remain isotropic and thermal.\footnote{One can also show this directly from the Vlasov equation \eqref{eq:kinetic} if we use equation \eqref{eq:Liouville} to write it as $\partial f/\partial t  =-  [f,H]$ where $[\, , \,]$ is a Poisson bracket \citep{BT}.  Thus, for the DF to evolve it must have gradients in phase space. A thermal distribution is by definition uniform in phase space at fixed $a$, and so never evolves at all.}

Of course, there is no reason to expect an arbitrary eccentricity DF to be of power law form. Indeed, random pairing of widely-separated stars formed in the turbulent interstellar medium gives rise to an eccentricity distribution significantly different from a power law form \citep{xu2023wide}. Thus, we will now broaden our definitions of subthermal and superthermal distributions somewhat. We will, however, demand that:
\begin{enumerate}
    \item $P(e) \to 0$ for $e \to 0$,
    \item $P(e)$ is monotonically increasing, i.e. $P'(e) > 0$,
    \item $P(e)$ is either concave or convex, i.e. $P''(e)$ never changes sign, and
    \item $P(e)$ does not diverge for any $e$.
\end{enumerate}
These restrictions are well-motivated in the context of wide binaries with $a\gtrsim 10^3$ AU in the Galactic field (see e.g. Figure 4 of \citealt{xu2023wide}). For instance, it is unlikely that any formation process will produce an abundance of nearly circular wide binaries: if a binary forms at a particular separation $\bm{r}$, there is only \textit{one} relative speed $v$ that will give rise to a circular orbit, while all others will produce an eccentric orbit.

Of course, there are scenarios in which
(i)-(iv) are not all satisfied.  In particular, (ii) and (iii) are invalid for 
wide binaries formed through the dissolution of unstable triples \citep{reipurth2012formation}.
Moreover, the DF of short-period binaries tends to build up at $e=0$ due to 
tidal dissipation \citep{price2018binary}. We will not consider such cases here.

Given the conditions (i)-(iv), we say that $P(e)$ is \textit{subthermal} if its slope at low eccentricity $P'(0) > 2$, and \textit{superthermal} if $P'(0) < 2$. An important consequence of conditions (ii) and (iii) is that for any allowed DF $P$, there is a unique nonzero eccentricity $\tilde{e}$ at which $P(\tilde{e}) = 2\tilde{e}$. We say a distribution $P_1$ is ``more subthermal (superthermal)'' than another distribution $P_2$ if both $P_1$ and $P_2$ are subthermal (superthermal) and the eccentricity $\tilde{e}_1$ at which $P_1(\tilde{e}_1) = 2\tilde{e}_1$ is less than (greater than) the eccentricity $\tilde{e}_2$ at which $P_2(\tilde{e}_2) = 2\tilde{e}_2$, i.e. $P_1$ has more of a surplus (deficit) of low $e$ (high $e$) binaries than $P_2$. This classification scheme is consistent with our definition of sub- and superthermal power law eccentricity DFs (see Figure \ref{fig:PFf_powerlaws}). Note that some of these constraints can be expressed more succinctly in terms of the angular momentum DF $F(j)$: conditions (i)-(iv) imply that $F(0) = 0$ always, and the superthermal DFs are simply those with $F(1) < 2$, while subthermal DFs are those with $F(1) > 2$.

\begin{figure*}
    \centering
    \includegraphics[width=0.99\textwidth]{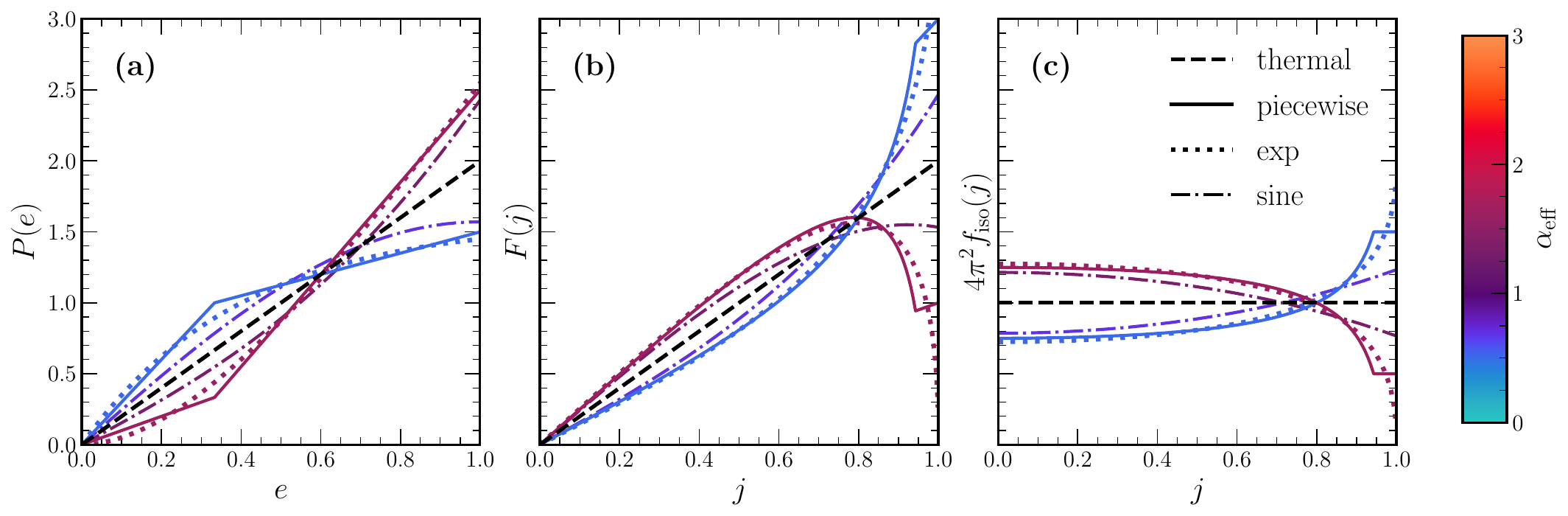}
    \caption{Some non-power law DFs that meet the criteria (i)-(iv) described in \S\ref{sec:subvssuper}. The piecewise subthermal DF (solid) is defined using equation \eqref{eq:piecewise_sub} with $\beta = 1/3$ and has $\alphaeff = 0.52$. The exponential subthermal DF (dotted) is defined using equation \eqref{eq:exp_sub} with $\gamma = 5/2$ and has $\alphaeff = 0.53$. The sine subthermal DF (dot-dashed) is defined using equation \eqref{eq:sine_sub} and has $\alphaeff = 0.70$. The corresponding superthermal DFs (same linestyles as their subthermal counterparts) are defined as reflections across the thermal distribution, and have $\alphaeff = 1.64$, $1.61$, and $1.34$ respectively. Similar to Figure \ref{fig:PFf_powerlaws}, panels (a) and (b) show the marginal eccentricity and angular momentum DFs, while panel (c) shows the resulting full DF $f_\mathrm{iso}$, assuming the binaries are oriented isotropically. In each panel, the black dashed line indicates the thermal distribution ($\alpha = 1$).}
    \label{fig:PFf_others}
\end{figure*}

In Figure \ref{fig:PFf_others} we give some examples of non-power law DFs that meet the criteria (i)-(iv). In particular, we define a ``piecewise'' subthermal distribution
\begin{align}
    \label{eq:piecewise_sub}
    P_\mathrm{piecewise,\ sub}(e; \beta) =
    \begin{cases}
        \frac{1}{\beta} e & e \leq \beta \\
        \frac{\beta}{(1-\beta)^2}(e-\beta)+1 & e > \beta,
    \end{cases}
\end{align}
a ``sine'' subthermal distribution
\begin{align}
    \label{eq:sine_sub}
    P_\mathrm{sine,\ sub}(e) = \frac{\pi}{2}\sin\left(\frac{\pi}{2}e\right),
\end{align}
and an ``exponential'' subthermal distribution
\begin{align}
    \label{eq:exp_sub}
    P_\mathrm{exp,\ sub}(e; \gamma) = \frac{\gamma}{\gamma-1+\exp(-\gamma)}(1-\exp(-\gamma e)),
\end{align}
with corresponding superthermal distributions defined as reflections across the thermal distribution,
\begin{align}
    \label{eq:corresponding_super}
    P_\mathrm{\_,\ super}\equiv 4e-P_\mathrm{\_,\ sub}.
\end{align}
The parameters $\beta$ and $\gamma$ set the shapes of the piecewise and exponential DFs respectively (e.g. the slope near $e = 0$ and $e = 1$); the plots in Figure \ref{fig:PFf_others} and throughout the remainder of this work use $\beta = 1/3$ and $\gamma = 5/2$. The colors in this Figure correspond to the ``effective power law index'' $\alpha_\mathrm{eff}$ of each DF. To define this effective index, we let $\tilde{e}$ be the eccentricity at which $P$ intersects the thermal distribution ($P(\tilde{e}) = 2\tilde{e}$), and then require $\int_0^{\tilde{e}} \md e\, P(e) = \int_0^{\tilde{e}} \md e\, P^{(\alpha_\mathrm{eff})}(e)$, i.e.
\begin{align}
\alpha_\mathrm{eff} = \frac{\log  \int_0^{\tilde{e}} \md e\, P(e)}{\log \tilde{e}} - 1.
    \label{eq:alphaeff}
\end{align}
It follows that $P$ has the same total surplus or deficit of low-$e$ binaries compared to the thermal DF as a genuine power law eccentricity DF $P^{(\alphaeff)}$ given by equation \eqref{eq:ecc_power_law}. The advantage of introducing $\alpha_\mathrm{eff}$ is that it allows us to classify a broad array of initial eccentricity DFs with a single parameter: analogous to power laws, an eccentricity DF is subthermal (superthermal) if it satisfies $\alphaeff < 1$ ($\alphaeff > 1$). 
Thus, $\alphaeff$ provides a useful measure of an eccentricity DF's deviation from thermality.\footnote{As an example, the analytic distribution given in equation (13) of \citet{xu2023wide} has an effective power law index of $\alphaeff \approx 1.43$.}
As we will see in \S\ref{sec:Milky_Way}, $\alphaeff$ also turns out to be a reliable predictor of the phase-mixed DF towards which an ensemble of wide binaries is driven by Galactic tides.


\subsection{Coarse-graining and phase-mixing}
\label{sec:coarse_graining}


In principle, equation \eqref{eq:kinetic} gives the exact continuum description of the DF at the most fine-grained possible level, assuming there are infinitely many binaries in our sample. If we could follow the DF with this perfect resolution, we would find that each infinitesimal piece of ``probability fluid'' retained the same density $f$ as it moved through phase space. Of course, in reality, our sample is always finite, and any measurement of the DF always involves some effective binning, which is equivalent to the \textit{mixing} of nearby phase space fluid elements \citep{Dehnen_2005}. This means that in practice, at some scale in phase space (potentially an extremely small scale, but finite nonetheless) the DF must be coarse-grained.

The result of coarse-graining is that Liouville's theorem no longer holds in the exact sense of every infinitesimally small phase-space fluid element conserving its value of $f$. Instead, it implies the following two \textit{local} properties (\citealt{THL_1986, Dehnen_2005}), which we will find useful for proving results about eccentricity DFs in the next subsection\footnote{These properties require phase space to be finite, which is true in our case at each for fixed semimajor axis $a$. For a visualization of the space, see Figure 1 of H22.}:
\begin{itemize}
    \item The minimum value of $f$ at time $t$, namely $f_\mathrm{min}(t)$, cannot decrease: $\md f_\mathrm{min} / \md t \geq 0$.
    \item The maximum value of $f$ at time $t$, namely $f_\mathrm{max}(t)$, cannot increase: $\md f_\mathrm{max} / \md t \leq 0$.
\end{itemize}
Coarse-graining also violates the exact conservation of $C_2$. That is, rather than being invariant, the $C_2$ value of the coarse-grained DF need only be non-increasing:
\begin{align}
    \frac{\md C_2}{\md t} \leq 0,
    \label{eq:C2_decrease}
\end{align}
which is directly analogous to the non-decreasing nature of entropy.

\begin{figure*}
    \centering
    \includegraphics[width=0.99\textwidth]{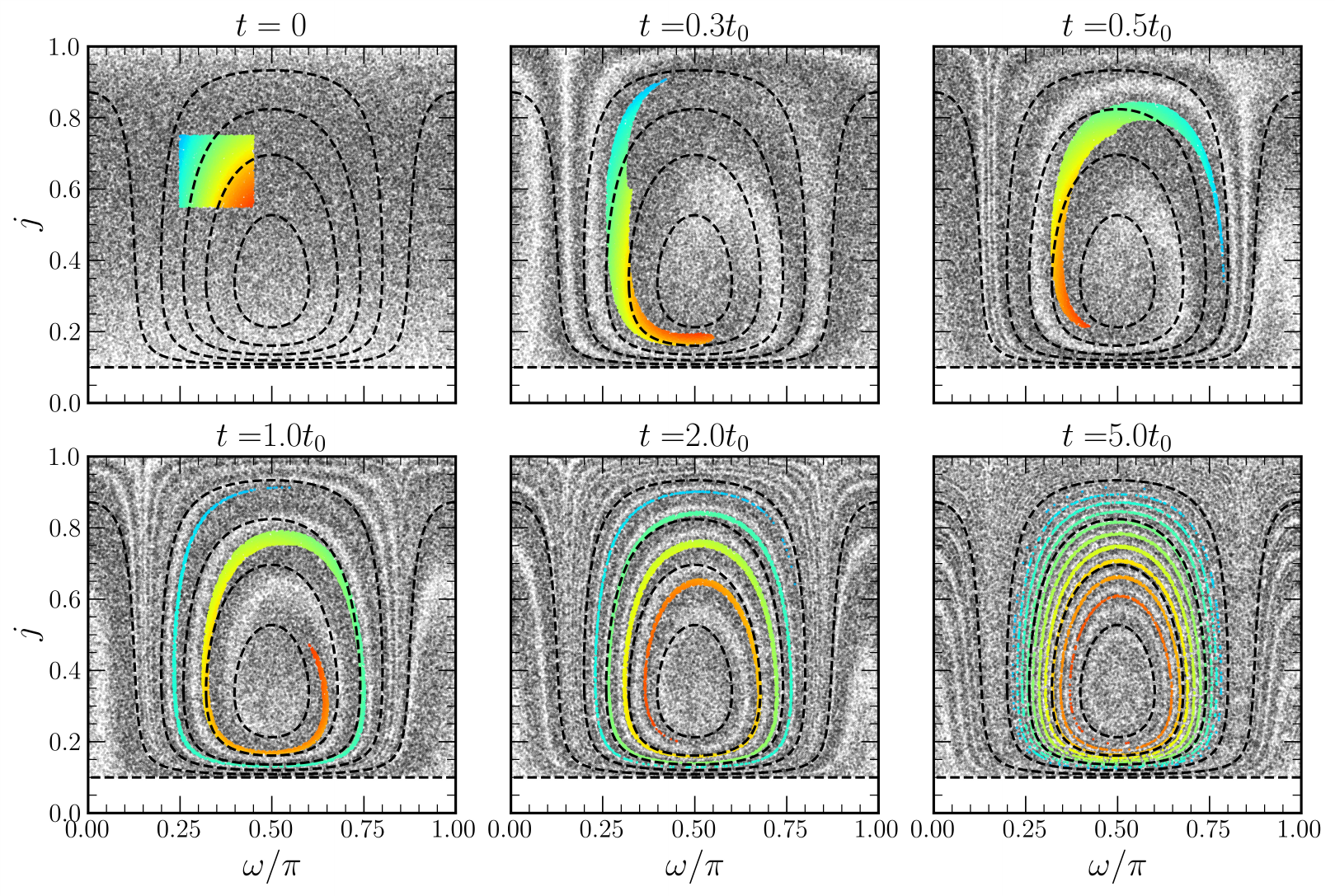}
    \caption{Snapshots of the DF in the  $(\omega, j)$ plane at fixed $j_z = 0.1$ for a simulation of wide binaries orbiting the Galactic disk (\S\ref{sec:Milky_Way}) initialized with an isotropic $\alpha = 2$ power law distribution. Each point represents a single binary ($10^5$ are shown in each panel), and black dashed curves are contours of constant $H_\Gamma$ as defined in equation \eqref{eq:def_HGamma} with $\Gamma=1/3$. The time unit $t_0$ is defined in equation \eqref{eq:t0} and corresponds to approximately five secular periods for orbits librating about the fixed point at $(\omega, j) = (\pi/2, 0.33)$. We color a subset of the binaries (those with initial conditions $(\omega/\pi, j)\in [0.25, 0.45]\times[0.55, 0.75]$) by their $H_\Gamma$ value to highlight the mixing process.}
    \label{fig:phasemixing}
\end{figure*}
If the Hamiltonian $H$ is time-independent (as it will be for our model in \S\ref{sec:Milky_Way}), then equation \eqref{eq:kinetic} tells us that, ignoring coarse-graining, each infinitesimal phase space fluid element traverses a contour of constant $H$, from which it never deviates. But because different $H$-contours tend to correspond to different phase space velocities, nearby phase space fluid elements diverge from each other, resulting in a shearing or \textit{phase-mixing} of the DF. As an example of this mixing process, in Figure \ref{fig:phasemixing} we show snapshots of the simulated phase space density in the $(\omega, j)$ plane for an ensemble of 100,000 wide binaries with an initially isotropic $\alpha = 2$ power law DF and fixed $j_z = 0.1$, using a particular time-independent Hamiltonian model for Galactic tides that we will introduce in \S\ref{sec:Milky_Way}. The dashed black curves in each panel correspond to iso-contours of this Hamiltonian along which individual binaries move, and the timescale $t_0$ given by equation \eqref{eq:t0} is roughly equal to five secular periods for orbits that librate around the fixed point at $(\omega, j) = (\pi/2, 0.33)$. We see that phase mixing leads to ever-finer structure in the DF. To illustrate this further, we highlight a subset of the binaries, namely those with initial conditions $(\omega/\pi, j)\in [0.25, 0.45]\times[0.55, 0.75]$, coloring them by their value of the Hamiltonian $H_\Gamma$. We see that by $t=5t_0$ these colored binaries are distributed almost evenly along their individual iso-Hamiltonian contours.

Eventually, the minimum scale of this structure is smaller than the scale over which we must coarse-grain the DF. By Jeans' theorem, the coarse-grained DF will then reach a steady state in which it depends only on $H$; this can be calculated by smearing the binaries of the initial DF $f_0$ uniformly over the $H$ contours on which they were initialized (H22):
\begin{align}
    \label{eq:finfty}
    f_\infty(\bw) = \left(\oint_{\gamma(\bw)} \md\lambda f_0(\bw'(\lambda))\right) \bigg/ \left( \oint_{\gamma(\bw)} \md\lambda \right).
\end{align}
Here $\gamma(\bw) = \{\bw' | H(\bw') = H(\bw)\}$ is a contour of constant $H$ parameterized by $\lambda$, the numerator measures the initial fraction of binaries present on the contour, and the denominator is the contour's length. However, the utility of equation \eqref{eq:finfty} is limited to time-independent $H$ and to late times when phase mixing is complete, whereas the other results of this subsection are much more general, relying on neither of these assumptions.


\subsection{Fundamental constraints on eccentricity DFs}
\label{sec:proofs}


For the remainder of this work, we will assume that the inner orbital planes of the binaries are initially oriented isotropically in space (see \S\ref{sec:discussion} for a discussion of this assumption). Then we may combine the results of the preceding subsections to prove the following four general statements about eccentricity DFs:
\begin{enumerate}
    \item Superthermal DFs cannot become more superthermal.
    \item Subthermal DFs cannot become more subthermal.
    \item Superthermal DFs cannot become subthermal.
    \item Subthermal DFs cannot become superthermal.
\end{enumerate}
These were conjectured by H22 in the particular context of wide binaries in the Galactic disk experiencing quadrupolar tides, and in the steady state (fully phase-mixed) limit given by equation \eqref{eq:finfty}. However, they are actually true for \textit{any} Keplerian ensemble whose exact evolution is governed by the Vlasov equation \eqref{eq:kinetic}, regardless of the particular form of $H$, and for \textit{any} time. The proofs are as follows.

\begin{enumerate}
    \item[\textbf{Proof of (i):}] An initially isotropic, superthermal DF $f_0(j)$ has its maximum at $j=0$ --- see Figures \ref{fig:PFf_powerlaws}c and \ref{fig:PFf_others}c for illustration. Now, suppose we wanted to make this DF more superthermal. If it remained isotropic, this would mean that $f(j=0)$ would have to increase, which contradicts $\md f_\mathrm{max} / \md t \leq 0$. Of course, under a generic Hamiltonian flow $H$, the DF will \textit{not} remain isotropic, but this simply means that at $j=0$, the DF is more concentrated at some ``orientations'' (values of $(\omega, \Omega, j_z)$) than others. Since the total value of the marginal DF $F(j)$ must be zero at $j=0$, and this marginal DF is found by integrating out the orientation dependence, there will inevitably be at least one location where $f(j=0)$ exceeds $f_0(j=0)$, again contradicting $\md f_\mathrm{max} / \md t \leq 0$. $\square$ \newline
    
    \item[\textbf{Proof of (ii):}] The same argument as in (i), but applied at $j=1$. $\square$ \newline
    
    \item[\textbf{Proof of (iii):}] If one allows only power law DFs (both initial and final), then this follows immediately from Figure \ref{fig:PFf_powerlaws}, along the lines of the argument in proof (i).  Precisely, all superthermal isotropic power laws have a maximum of $f_0$ at $j=0$, and this maximum is finite. On the contrary, all subthermal isotropic power laws' $f_0(j)$ diverge as $j\to 1$.  Thus, to convert a superthermal DF to a subthermal DF would be to increase the maximum $f$, contravening $\md f_\mathrm{max} / \md t \leq 0$.
    
    A very similar argument applies to non-power law DFs (see Figure \ref{fig:PFf_others}), although we have to very slightly toughen the restrictions on our definitions of subthermal (superthermal) DFs $P(e)$ to those with negative (positive) curvature at the origin, $P''(0) < 0$ ($P''(0) > 0$). Strictly, this excludes the piecewise-linear DFs and sine-DFs that we discussed in \S\ref{sec:distribution_functions}; these DFs are problematic for our argument since their $f_\mathrm{iso}(j)$ does not diverge as $j\to 1$. Nevertheless, since these DFs can be made to fit our new criteria with only very minor alterations (e.g. by using an exponential model with the same $\alphaeff$ instead of the piecewise form, as shown in Figure \ref{fig:PFf_others}), we consider case (iii) proven. $\square$ \newline
    
    \item[\textbf{Proof of (iv):}] Similar to (iii), except this time the argument is that one cannot decrease the minimum $f$. For power laws it is clear from Figure \ref{fig:PFf_powerlaws} that going from a subthermal to superthermal DF is not allowed because subthermal DFs have finite $f_0$ everywhere whereas superthermal DFs have $f_0(1) = 0$. The same is true for non-power law DFs if we make the same additional restrictions we did in (iii), namely subthermal (superthermal) DFs $P(e)$ must have $P''(0) < 0$ ($P''(0) > 0$). $\square$
\end{enumerate}

Less formally, statements (i)-(iv) agree with the intuition gleaned from considering the Casimir $C_2$, which we recall measures the nonuniformity of the DF. Indeed, if we restrict to only initial and final power law DFs, then statements (i) and (ii) are already apparent from Figure \ref{fig:C_alphas}. To see this, define $\overline{f}$ as the average of $f$ over $\omega$, $\Omega$, and $j_z$, i.e. 
\begin{align}
    \label{eq:f_avg_C2}
    \overline{f} = \frac{F(j)}{8\pi^2j}.
\end{align}
For an initially isotropic distribution, then, $\delta f = 0$ and $C_2 = \overline{C}_2$. Tides necessarily induce anisotropy (see \S\ref{sec:results_inclination}), so $\delta f^2$ (and hence $\delta C_2$) must be positive at later times. Since the total $C_2 = \overline{C}_2 + \delta C_2$ is conserved under exact Hamiltonian evolution (and must in fact decrease under coarse graining) the value of $\overline{C}_2$ must necessarily decrease. This corresponds to an eccentricity distribution evolving toward the thermal minimum at $\alpha = 1$.


\section{Wide binaries in the Solar Neighborhood}
\label{sec:Milky_Way}


The results of \S\ref{sec:ensemble_evolution} apply to an arbitrary ensemble of Keplerian orbits evolving under some generic, weak tidal Hamiltonian $H$. In this section, we will focus on the particular case of wide binaries orbiting in the Galactic disk. In \S\ref{sec:Galactic_Disk} we introduce our simple model for the binary-disk interaction. In \S\ref{sec:numerical_examples}, we present numerical examples of wide binary ensembles evolving under the Galactic tides. These examples serve to illustrate the claims made in \S\ref{sec:ensemble_evolution}, and also provide more insight into the specific problem of wide binaries in the Galaxy than can be deduced on general grounds.


\subsection{Interaction of a single wide binary with the Galactic tide}
\label{sec:Galactic_Disk}


Let $\Phi$ be the potential of the Galactic disk. The Hamiltonian describing the tidal perturbation that this disk exerts upon a wide binary can always be expanded in terms of a small parameter $\sim a/h$, where $h\sim 200$\,pc is the disk scale height. The lowest order contribution to this Hamiltonian is the quadrupolar term; octupolar and higher order terms are smaller by $\mathcal{O}(a/h) \sim 10^{-3}$, and so are negligible. After averaging over the inner orbit (i.e. over the mean anomaly $\eta$), the perturbing Hamiltonian reads \citep{Hamilton2019-jn}:
\begin{align}
    \label{eq:H1SA}
    \delta H(\bm{\theta}, \bm{I},t) = \frac{1}{2}\sum_{\ell,k} \Phi_{\ell k} \langle r_\ell r_k \rangle_\eta.
\end{align}
Here, $r_\ell$ and $r_k$ are components of the relative separation vector $\bm{r}=(x,y,z)$ between the stars in the binary, and $\Phi_{\ell k} \equiv [\partial^2\Phi/ \partial r_\ell \partial r_k ]_{\bm{R}_\mathrm{g}(t)}$ is the tidal tensor evaluated at the binary's current barycentric position in the Galaxy $\bm{R}_\mathrm{g}(t)$. The $\eta$-averaged quantities $\langle r_\ell r_k \rangle_\eta$ are given explicitly in Appendix A of \cite{Hamilton2019-jn}.

If we also average $\delta H$ over the outer orbit, then we get the ``doubly-averaged'' Hamiltonian, which is equivalent to equation \eqref{eq:H1SA} with $\Phi_{\ell k}$ replaced with its time-averaged value $\overline{\Phi}_{\ell k}$. The result is
\begin{align}
    \label{eq:def_H}
    \delta H = \frac{Aa^2}{8} H_\Gamma,
\end{align}
where $A \equiv \overline{\Phi}_{xx} + \overline{\Phi}_{zz}$ measures the strength of the tides, and
\begin{align}
    \label{eq:def_HGamma}
    H_\Gamma \equiv \frac{1}{j^2}\big[(j^2 -  3\Gamma  j_z^2)( 5-3j^2) -15\Gamma(j^2-j_z^2)(1-j^2) \cos 2\omega \big]
\end{align}
is a dimensionless Hamiltonian, parameterized by $\Gamma \equiv [\overline{\Phi}_{zz} - \overline{\Phi}_{xx} ]/(3A)$. Naively, this doubly-averaged Hamiltonian is only appropriate if the tidal perturbation is sufficiently weak, so that e.g. the timescale for evolution of eccentricity is long compared to the outer orbital period \citep{Hamilton2019-zl}. However, for wide binaries in the Solar neighborhood, the equations of motion derived from the Hamiltonian \eqref{eq:def_H} are usually accurate regardless of whether this is true. The reason is that if the outer orbit of the binary is close to epicyclic, then $\Phi_{\ell k}$ is already approximately time-independent (see Appendix C of \citealt{Hamilton2019-jn} and \citealt{Heisler1986-eb}), so the singly- and doubly-averaged Hamiltonians are identical.

An even greater simplification, which we will use throughout the remainder of our analysis, follows from the fact that the tidal tensor in the Galactic disk is dominated by $\Phi_{zz}$, so that $\Gamma \approx 1/3$ and $A \approx \overline{\Phi}_{zz} \approx 4\pi G \rho_0$ is the square of the local vertical epicyclic frequency (where $\rho_0$ the local mass density). Note also that $H_\Gamma$ is independent of $\Omega$, so by Hamilton's equations \eqref{eq:Hamiltons_Equations}, $J_z$ (and hence $j_z$) is constant. Thus, each binary is represented by a point $\bw$ (see equation \eqref{eq:def_DF}) in phase space, and binaries evolve on contours of constant $H_\Gamma$ in planes of $(\omega, j)$ at fixed $j_z$ --- see Figure \ref{fig:phasemixing}. The nodal angle $\Omega$ also evolves under secular dynamics according to $\md\Omega/\md t \propto \partial H_\Gamma/\partial j_z$, but it is effectively decoupled from the rest of the phase space\footnote{H22 erroneously claimed that an initially uniform distribution in $\Omega$ would always remain uniform. This is not true, but does not affect any of the conclusions of that paper or the present one.}, so can be ignored here.

\begin{figure}
    \centering
    \includegraphics[width=0.45\textwidth]{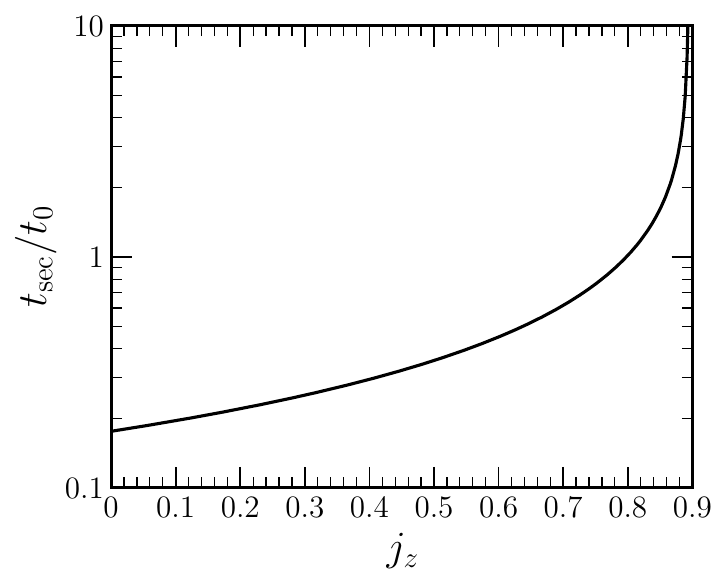}
    \caption{The secular period $t_\mathrm{sec}$ for oscillations around the fixed point at $\omega=\pi/2$ (e.g. Figure \ref{fig:phasemixing}), as a function of $j_z$.
    (There is no fixed point for $j_z > \sqrt{4/5} \approx 0.89$ --- see equation (13) of \citealt{Hamilton2019-jn}).
    The timescale $t_0$ is defined in equation \eqref{eq:t0}.}
    \label{fig:tsec_vs_jz}
\end{figure}

The secular period $t_\mathrm{sec}$ --- i.e. the time it takes for the binary to perform one full oscillation in the $(\omega, j)$ plane --- is given in equation (33) of \cite{Hamilton2019-zl}. At fixed $j_z$, the precise value of $t_\mathrm{sec}$ depends on the contour $H_\Gamma$ to which the binary belongs, but a reliable benchmark value is provided by the value of $t_\mathrm{sec}$ at the fixed point at $\omega = \pi/2$ (equation (12) of \citealt{Hamilton2019-jn}). In Figure \ref{fig:tsec_vs_jz} we plot this secular period as a function of $j_z$, in units of 
\begin{align}
    \label{eq:t0}
    t_0 & \equiv \frac{8L}{Aa^2} \approx \frac{\pi}{4}\frac{T_Z^2}{T_b} \\
    & \approx 4 \,\mathrm{Gyr}\,\left( \frac{\rho_0}{0.2 \Msun \,\mathrm{pc}^{-3}}\right)^{-1}
    \left( \frac{m_b}{\Msun} \right)^{1/2} \left(\frac{a}{10^4\,\mathrm{AU}}\right)^{-3/2},
\end{align}
where $T_Z = 2\pi/\sqrt{4\pi G\rho_0}$ is the period of vertical oscillations of the binary's outer orbit in the Galactic potential, and $T_b = 2\pi/n$ is its inner orbital period. We see that for the majority of $j_z$ values, $t_\mathrm{sec}$ is comparable to (but smaller than) $t_0$, justifying the estimate \eqref{eq:t_sec_numerical}. This suggests that wide binaries with $a\sim 10^4$ AU will have typically have completed a few secular oscillations over the lifetime of the Galaxy, but those with $a\sim 10^3$ AU will not have completed even one such oscillation.


\subsection{Numerical examples}
\label{sec:numerical_examples}


Next, we carry out numerical simulations of ensembles of binaries evolving under the Hamiltonian \eqref{eq:def_HGamma} with $\Gamma = 1/3$. The equations of motion for $\bw$ are given by differentiating this Hamiltonian according to \eqref{eq:Hamiltons_Equations}. We integrate the equations of motion forward in time using the DOP853 method \citep{DOP853} implemented in scipy \citep{scipy}. Note that all binaries in the ensemble are independent of each other, i.e. there are no binary-binary interactions. In each simulation we draw $N=10^5$ binaries randomly from an initial DF which is isotropic in angles (uniform in $j_z/j = \cos i$, $\omega$, and $\Omega$).

We consider several different initial eccentricity distributions:
\begin{itemize}
    \item power law distributions with $\alpha = 0$ (subthermal), $\alpha = 1$ (thermal), and $\alpha = 2$ (superthermal);
    \item a ``piecewise'' subthermal distribution (see equation \eqref{eq:piecewise_sub}) with $\beta = 1/3$, and its superthermal counterpart;
    \item a ``sine'' subthermal distribution (see equation \eqref{eq:sine_sub}) and its superthermal counterpart;
    \item an ``exponential'' subthermal distribution (see equation \eqref{eq:exp_sub}) with $\gamma = 5/2$ and its superthermal counterpart.
\end{itemize}
These choices of $\alpha$, $\beta$, and $\gamma$ allow us to investigate a range of DFs which have varying deviations from thermality, with the aim of understanding the origin of the observed power law $\alpha\approx 1.3$. Additionally, with these choices of $\beta$ and $\gamma$ for the piecewise and exponential distributions respectively, we are able to explore the evolution of ensembles that have similar effective power law indices $\alphaeff$ (see \S\ref{sec:subvssuper}) despite having differing functional forms. We summarize this information in Table \ref{tab:sim_summary}, where we also give the corresponding initial and measured final power law indices of the eccentricity DF.

\begin{table}
    \centering
    \begin{tabular}{c|c|c|c} \hline
        Name & Initial $P(e)$ & $\alpha_\mathrm{i}$ & $\alpha_\mathrm{f}$ \\ \hline
        power law $\alpha=0$ & \eqref{eq:ecc_power_law} with $\alpha = 0$ & 0 & 0.35 \\
        power law $\alpha = 2$ & \eqref{eq:ecc_power_law} with $\alpha = 2$ & 2 & 1.47 \\
        piecewise subthermal & \eqref{eq:piecewise_sub} with $\beta = 1/3$ & 0.52 & 0.78 \\
        piecewise superthermal & \eqref{eq:piecewise_sub} and \eqref{eq:corresponding_super} with $\beta = 1/3$ & 1.64 & 1.31 \\
        exp subthermal & \eqref{eq:exp_sub} with $\gamma = 5/2$ & 0.53 & 0.76 \\
        exp superthermal & \eqref{eq:exp_sub} and \eqref{eq:corresponding_super} with $\gamma = 5/2$ & 1.61 & 1.31 \\
        sine subthermal & \eqref{eq:sine_sub} & 0.70 & 0.87 \\
        sine superthermal & \eqref{eq:sine_sub} and \eqref{eq:corresponding_super} & 1.34 & 1.14 
        \\ \hline
    \end{tabular}
    \caption{Summary of the initial eccentricity DFs used in our numerical examples, alongside the (effective) power law index of the initial eccentricity distribution as calculated using equation \eqref{eq:alphaeff}, and the best-fit power law index of the final, fully-mixed eccentricity distribution. The final power law indices are measured at $t = 10t_0$, well after the distributions have converged to steady state.}
    \label{tab:sim_summary}
\end{table}

An example of this integration was already given in Figure \ref{fig:phasemixing}, which showed the phase space density of $10^5$ binaries in the $(\omega, j)$ plane for fixed $j_z=0.1$, at different times $t$, for the initial $\alpha=2$ superthermal distribution. Recall that the black dashed lines are contours of constant $H_\Gamma$, along which the individual binaries are advected. As binaries on adjacent contours undergo secular oscillations at differing frequencies, the DF phase-mixes until binaries are spread uniformly over each contour. To quantify the effects of this mixing process, we measure the eccentricity and inclination distributions of each of the ensembles as they evolve.


\subsubsection{Initial vs. final eccentricity distributions}
\label{sec:results_eccentricity}


\begin{figure*}
    \centering
    \includegraphics[width=0.912\textwidth]{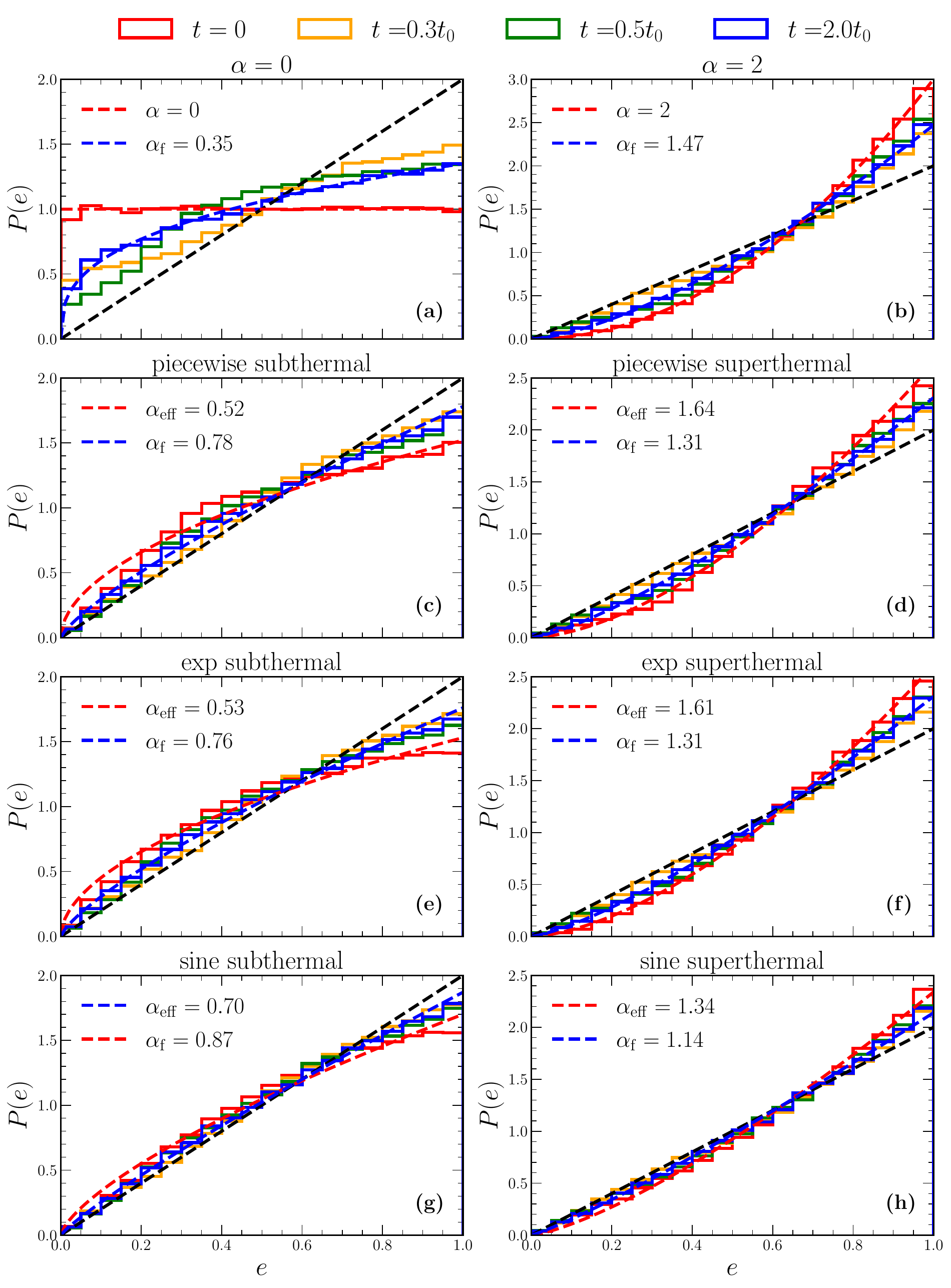}
    \caption{Evolution of the eccentricity DFs listed in 
    Table \ref{tab:sim_summary}. Different colored histograms represent snapshots at different times, and in each panel we overplot the initial distribution's corresponding ``effective'' power law distribution (red), a power law fit to the final distribution (blue), and the thermal distribution (black). DFs in the left column are subthermal and evolve toward an increased $\alpha_\mathrm{f}$, while those in the right column are superthermal and evolve toward a decreased $\alpha_\mathrm{f}$. Ensembles are also ordered with ``corresponding'' functional forms in the same row, and from farther from thermal $\alphaeff$ (top) to closer to thermal $\alphaeff$ (bottom).}
    \label{fig:snapshots_all}
\end{figure*}

In Figure \ref{fig:snapshots_all} we plot the eccentricity distribution $P(e, t)$ at several different times $t$ for each of the ensembles listed in Table \ref{tab:sim_summary}. In each set of simulations, the initial distribution (red) evolves toward a phase-mixed steady state (blue). Remarkably, regardless of the initial DF, we find that the steady state DF is always well-fit by a power law, whose index we call $\alpha_\mathrm{f}$.

For the initial power law distributions (panels (a) and (b)), the final power law index matches the predicted values from the semi-analytic calculation of H22, based on equation \eqref{eq:finfty}. Namely, the $\alpha = 0$ subthermal DF transforms to another (less) subthermal DF with $\alpha_\mathrm{f} = 0.35$, while the superthermal DF with $\alpha = 2$ transforms to a (less) superthermal DF with $\alpha_\mathrm{f} = 1.47$.

Turning to the initally non-power law DFs (panels (c)-(h)), we see that the final DF always meets the criterion to be considered ``closer to thermal'' specified in \S\ref{sec:subvssuper}. That is, characterizing each initial distribution with the effective power law index $\alphaeff$ defined by equation \eqref{eq:alphaeff} (red), the final index $\alpha_\mathrm{f}$ always lies in-between $\alphaeff$ and unity --- the distribution evolves closer to thermal, but always falls short of it. Moreover, we see that distributions with similar $\alphaeff$ all evolve toward very similar $\alpha_\mathrm{f}$ compare e.g. panels (c) and (d) with panels (e) and (f). Thus, the evolution of an ensemble's eccentricity distribution is determined primarily by its initial deficit or surplus of low-eccentricity binaries relative to the thermal distribution. These numerical results are all consistent with the fundamental constraints proven in \S\ref{sec:proofs}.\footnote{In fact our numerical results here somewhat extend the formal claims of \S\ref{sec:proofs}, since we recall that e.g. the $\alpha = 0$, piecewise, and sine DFs did not fit our criteria for acceptable DFs in those proofs.}

\begin{figure*}
    \centering
    \includegraphics[width=0.65\textwidth]{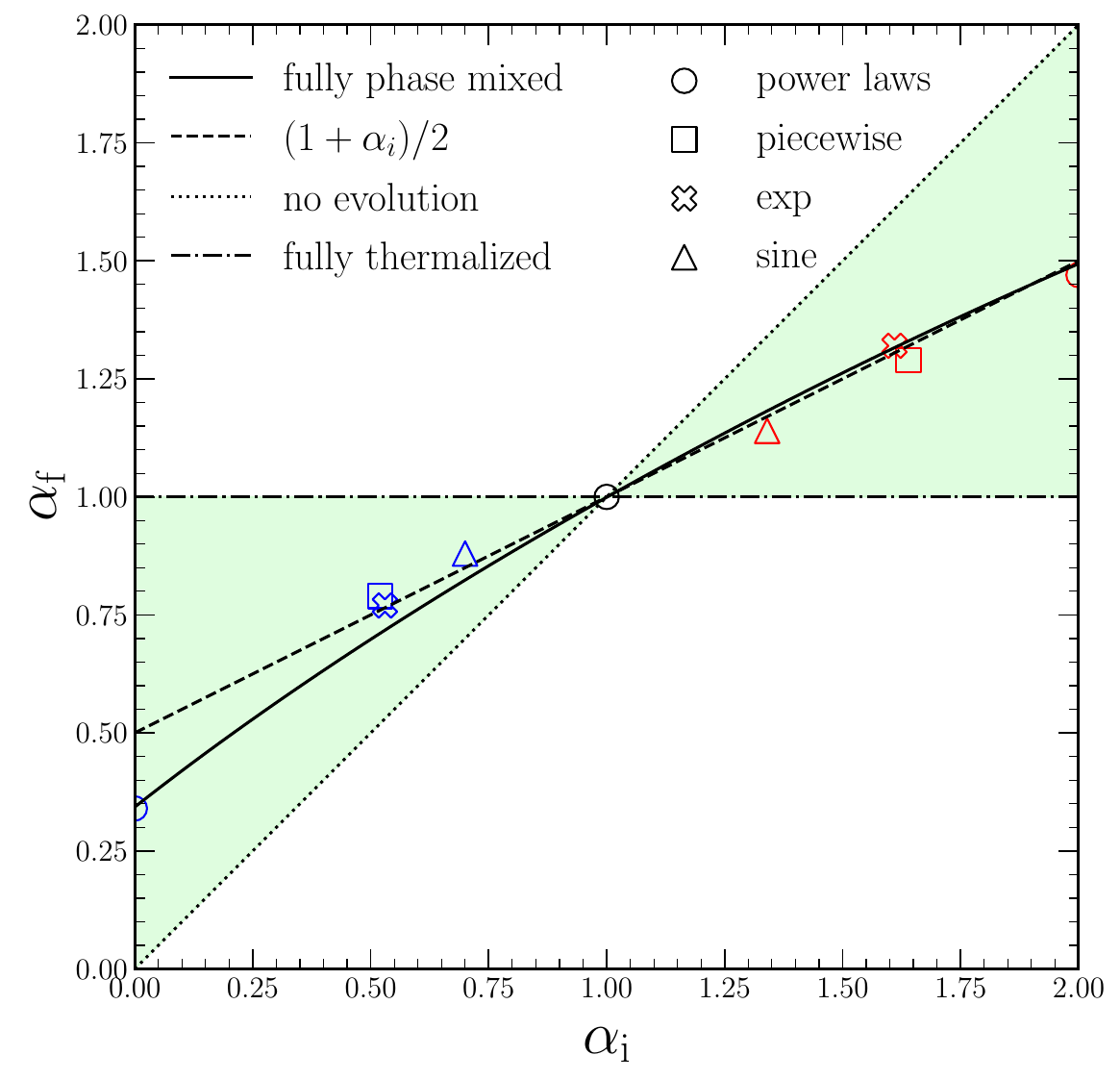}
    \caption{The final eccentricity distribution's best-fit power law index $\alpha_\mathrm{f}$ plotted against the initial power law index $\alpha_\mathrm{i}$ (taken to be the effective $\alphaeff$ for initially non-power law distributions). The general results proven in \S\ref{sec:proofs} require each ensemble to fall in the green region between the dotted (no evolution) and dot-dashed (fully thermalized) lines. The solid black curve is the fully phase-mixed distribution of equation \eqref{eq:finfty}, and the approximation $\alpha_\mathrm{f} = (1+\alpha_\mathrm{i})/2$ is overplotted as a dashed line.}
    \label{fig:alpha_f_vs_i}
\end{figure*}

To further demonstrate the utility of the effective power law index $\alphaeff$, we present Figure \ref{fig:alpha_f_vs_i}, in which we show the best-fit index of the final DF $\alpha_\mathrm{f}$ as a function of the initial index $\alpha_\mathrm{i}$. For the initially non-power law distributions we take $\alpha_\mathrm{i} = \alphaeff$ as calculated from equation \eqref{eq:alphaeff}. The solid black curve corresponds to the result for initial power law DFs according to equation \eqref{eq:finfty}. The circles show the numerical results for initial power law DFs, while the other symbols illustrate the results from the non power-law ensembles. The green shaded region in this plot corresponds to values of $\alpha_\mathrm{f}$ between $\alpha_\mathrm{i}$ and unity, which is the only allowed region according to the fundamental constraints proven in \S\ref{sec:proofs}.

We see that for power law indices not too far from unity (the observationally important regime, see \citealt{hwang2022wide}), the numerical results are well-approximated by the fitting formula
\begin{align}
    \label{eq:alphaf_approx}
    \alpha_\mathrm{f} \approx \frac{1}{2}(1 + \alpha_\mathrm{i}).
\end{align}
In other words, Galactic tides take an initial DF and drive it approximately ``halfway'' towards the thermal eccentricity distribution. Equations \eqref{eq:alphaeff}, \eqref{eq:t0}, and Figure \ref{fig:alpha_f_vs_i} therefore provide a complete ``forward model'' for (initially isotropic) binary eccentricity distributions. First, we use the given $P(e)$ to calculate $\alphaeff$. If the secular timescale is small enough that we expect the ensemble to be fully mixed, then we can extract the exact $\alpha_\mathrm{f}$ using the solid black curve in the Figure, and otherwise, we constrain the possible values of $\alpha_\mathrm{f}$ using the extent of the green region in the Figure.


\subsubsection{Initial vs. final inclination distributions}
\label{sec:results_inclination}


\begin{figure*}
    \centering
    \includegraphics[width=0.9\textwidth]{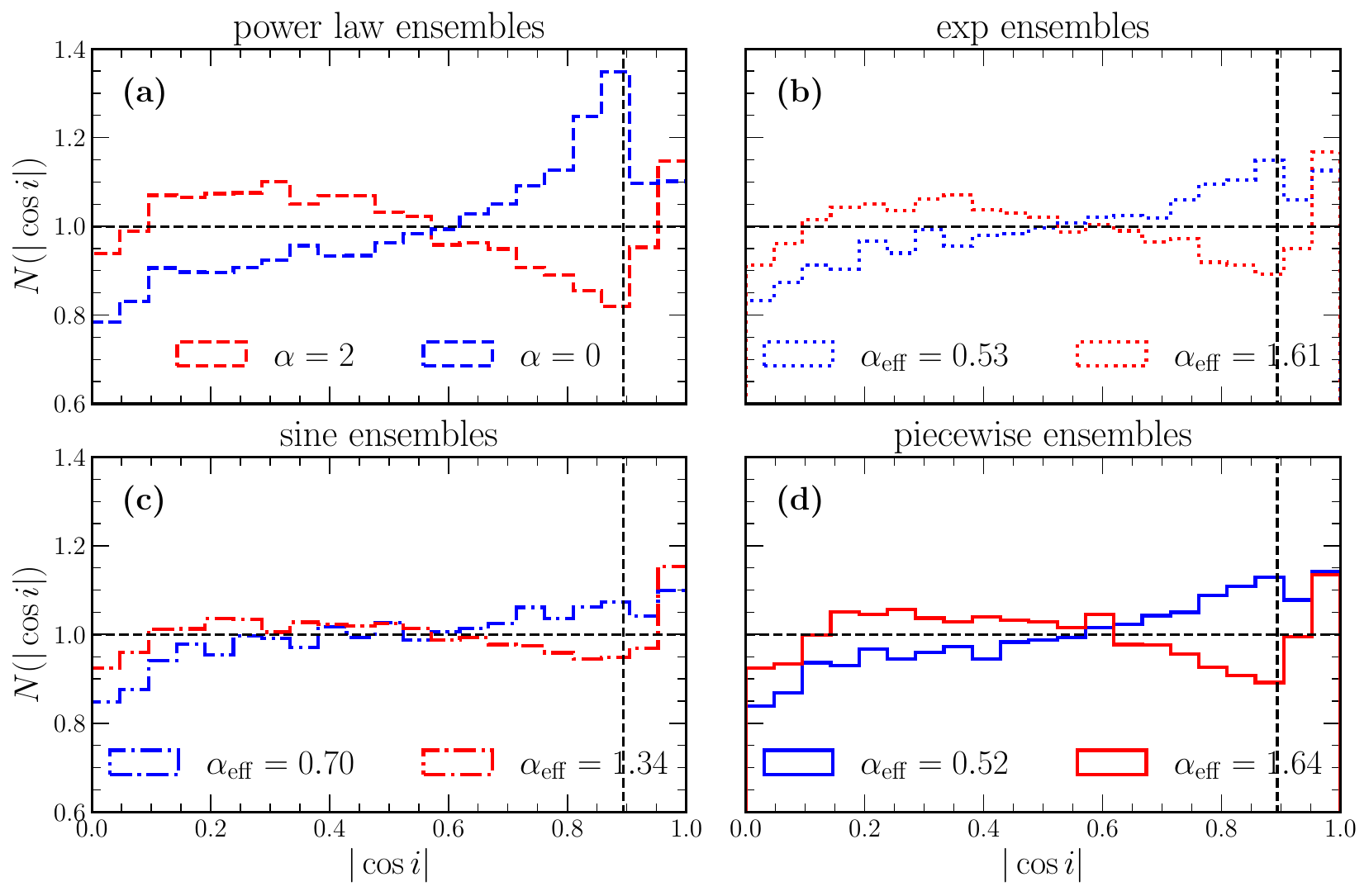}
    \caption{Inclination distributions for each ensemble in Table \ref{tab:sim_summary}. In each panel, the blue histogram represents the subthermal distribution of the functional form indicated in the title, the red histogram represents the corresponding superthermal distribution, and the initial (effective) $\alpha$ values are indicated in the legend. The horizontal black dashed line indicates the initial uniform distribution, and the vertical black dashed line indicates $\cos i \approx 0.89$, the predicted position of peaks and troughs from H22.}
    \label{fig:cosi_final_distributions}
\end{figure*}

We always choose our initial DF to be isotropic, i.e. uniform in $\cos i$. Evolution under Galactic tides does not preserve this isotropy; in Figure \ref{fig:cosi_final_distributions} we show the final, steady state inclination distributions $N(|\cos i|)$ for each ensemble (c.f. Figure 3 of H22)\footnote{Because the Hamiltonian \eqref{eq:def_HGamma} is an even function of $j_z$, the distribution of inclinations is also even, so we show the distribution of inclination \textit{magnitudes} here, which differs from the definition in equation \eqref{eq:marginal_cosi} by a factor of 2.}. We see that tides drive isotropic superthermal distributions (red) toward distributions with slight deficits at high $|\cos i|$, while the isotropic subthermal distributions do the opposite. As explained in H22, the peaks and troughs of these final inclination DFs are always located at $|\cos i| \approx 0.89$. The extent of the anisotropy present in the final phase-mixed DF is correlated with the initial deviation of the eccentricity DF from thermality. However, even the rather extreme DFs with $\alpha = 0$ and $\alpha = 2$ exhibit a maximum deviation from isotropy of $\lesssim 40\%$; the other examples shown here deviate from isotropy by at most $20\%$.


\subsubsection{Time-dependence}
\label{sec:results_time_dependence}


One key drawback of H22's calculation based on equation \eqref{eq:finfty} was the lack of information it gave about the evolution between the initial and final DFs, and the precise timescale over which the final DF is achieved. On the contrary, our numerical simulations allow us to investigate these questions in detail.

First, we know from Figure \ref{fig:phasemixing} that at early times, the distribution of binaries along each contour of constant $H_\Gamma$ is distributed much more unevenly than it will be at later times. Thus, evolution over the first secular oscillation or so (top row of Figure \ref{fig:phasemixing}) produces highly anisotropic, transient overdensities in the DF in each $(\omega, j)$ plane. Correspondingly, in Figure \ref{fig:snapshots_all} we see that the most dramatic evolution of the eccentricity DF $P(e,t)$ occurs at the earliest times. In fact, between $t=0$ and $t \sim  0.3t_0$, most of the $P(e, t)$ curves actually ``overshoot'' their final state (so the orange line in these panels comes closer to the thermal DF than does the final, blue one). Of course, this overshoot is never so dramatic as to cause the DF to cross from sub- to superthermal or vice versa, consistent with the fundamental constraints derived in \S\ref{sec:proofs}.

After $t\sim t_0$ (bottom row of Figure \ref{fig:phasemixing}), when most binaries have undergone multiple secular oscillations, the transient phase ends, and the ever finer-grained structure that is produced for $t>t_0$ does not substantially change the coarse-grained eccentricity DF. This is confirmed in Figure \ref{fig:snapshots_all}, which teaches us that the eccentricity DFs converge to within a few percent of their final phase-mixed DF state after $t\sim t_0\approx 4\,$Gyr $(a/10^4\mathrm{AU})^{-3/2}$. It follows that if Galactic tides were the only dynamical perturbation, binaries with $a\gtrsim 10^4$ AU would be well phase-mixed within the lifetime of the Galaxy.

\begin{figure}
    \centering
    \includegraphics[width=0.48\textwidth]{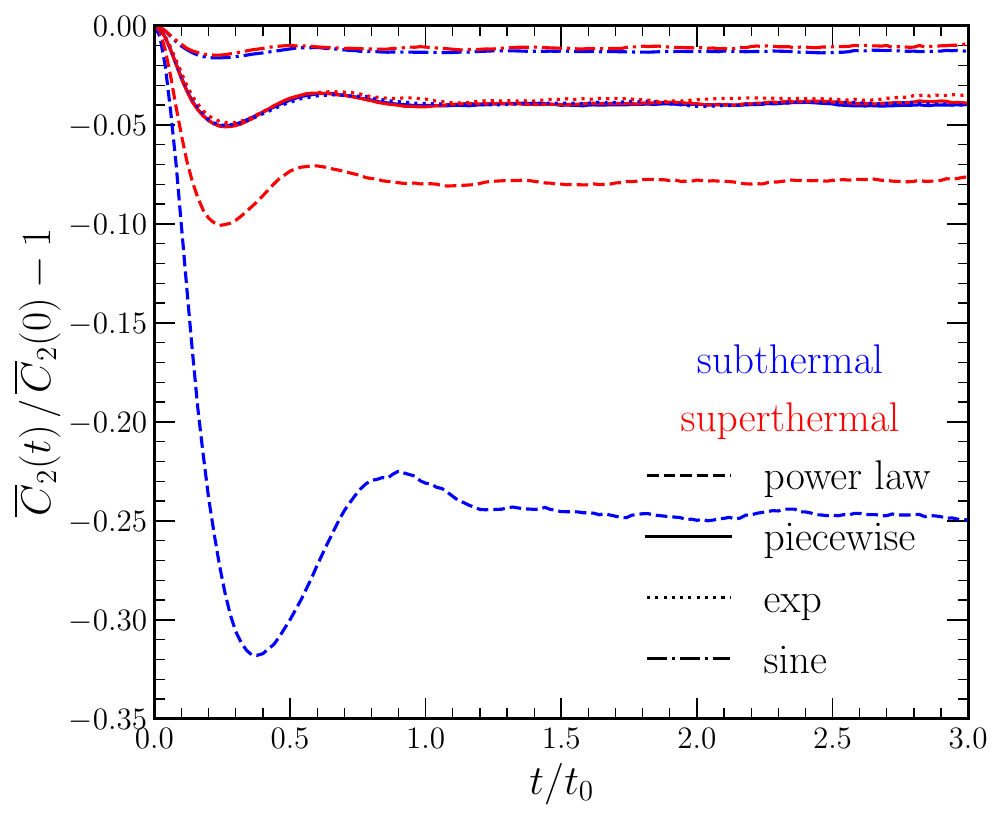}
    \caption{The fractional change in the orientation-averaged Casimir $\overline{C}_2$ as a function of time for the initial power law (dashed), piecewise linear (solid), exponential (dotted), and sine (dot-dashed) eccentricity distributions (see Table \ref{tab:sim_summary}). For each functional form, the red curve is the initially superthermal distribution, while the blue curve is the initially subthermal distribution.}
    \label{fig:C2_vs_t}
\end{figure}

To make this claim more precise, we study the evolution of the quadratic Casimir $C_2(t)$ given by equation \eqref{eq:Casimir}, which we split into an orientation-averaged part $\overline{C}_2$ that only depends on $j$ (see equation \eqref{eq:f_avg_C2}) plus an orientation-dependent fluctuation $\delta C_2$ which is initially zero. In the Appendix we discuss the method used to estimate these quantities from the numerical simulations. In Figure \ref{fig:C2_vs_t} we plot the fractional change in $\overline{C}_2$ as a function of time for each simulated ensemble. We observe a significant initial decrease in $\overline{C}_2$ (and hence increase in $\delta C_2$) as phase mixing drives the large overdensities shown in the upper row of Figure \ref{fig:phasemixing}. The $\overline{C}_2$ oscillations following this initial decrease are analogous to the entropy production fluctuations due to mixing observed in e.g. Figure 2 of \cite{BeraldoeSilva_entropy2017}. After $t\sim t_0$, $\overline{C}_2$ approaches a constant value across all ensembles, confirming that phase mixing is approximately complete by this time. As anticipated, the largest overall decreases in $\overline{C}_2$ are in the ensembles initialized farthest from the thermal DF ($\alpha_\mathrm{i}$ furthest from 1). Because the total $C_2$ of each ensemble is conserved (prior to coarse-graining), this is consistent with those ensembles producing the most anisotropic final distributions (\S\ref{sec:results_inclination}). Finally, we highlight that different ensembles with similar initial $\alphaeff$ exhibit very similar $\overline{C}_2$ behavior, suggesting that the $\alpha_\mathrm{eff}$ diagnostic is useful at all times, not only for fitting the final eccentricity DF.


\section{Discussion}
\label{sec:discussion}


In this section, we  discuss some of the caveats and limitations of our analysis, as well as some of the implications of our results for wide stellar binaries and other Keplerian systems.

The major piece of physics which we have left out of our study, which is essential to include if we are to draw strong astrophysical conclusions, is the effect of scattering by passing stars, gas clouds, dark matter substructure, and other perturbers in the Galaxy. We defer a full discussion of these effects to a companion paper (Paper II). There, we argue that the predominant scattering effect is from weak, impulsive, penetrative encounters by passing stars, and we investigate in detail the effect this has on the combined eccentricity and semimajor axis DF. Here, we limit ourselves to a brief assessment of the relevant timescales.

Impulsive encounters produce a systematic drift in binary semimajor axis; the tendency is for soft binaries to become gradually softer until they are eventually unbound (``ionized,'' see \citealt{Heggie1975}). For diffusive encounters with stars of mass $m_\mathrm{per}$, velocity dispersion $\sigma$ and number density $n$, the typical ionization timescale is given by \citep{BT}:
\begin{align}
    \label{eq:ionize}
    t_\mathrm{ion} & \sim \frac{0.02\sigma m_b}{G m_\mathrm{per}^2 n a \log \Lambda} \\
    & \approx 4\,\mathrm{Gyr}\,\left(\frac{\sigma}{40 \, \mathrm{km/s}}\right)\left(\frac{m_b}{M_\odot}\right)\left(\frac{m_\mathrm{per}}{M_\odot}\right)^{-2} \nn \\
    & \ \ \ \ \ \ \ \ \ \ \ \ \ \ \ \ \ \ \ \ \times \left(\frac{n}{0.2 \, \mathrm{pc}^{-3}}\right)^{-1}\left( \frac{a}{10^4 \mathrm{AU}}\right)^{-1}\left(\frac{\log \Lambda}{5} \right)^{-1},
\end{align}
where $\Lambda \sim a\sigma^2/(Gm_\mathrm{per})$. As we show in Paper II, the typical timescale for scattering to produce an $\mathcal{O}(1)$ change in a binary's \textit{eccentricity} is also $\sim t_\mathrm{ion}$. Let us therefore compare the ``scattering'' timescale $t_\mathrm{ion}$ with the secular timescale $\tsec$ over which Galactic tides are able to modify the eccentricity distribution.  Using equations \eqref{eq:t_sec_numerical} and \eqref{eq:ionize} and assuming  $\rho_0 \sim nm_\mathrm{per}$, the ratio is 
\begin{align}
    \frac{t_\mathrm{ion}}{t_\mathrm{sec}} \sim 4 \left(\frac{\sigma}{40 \, \mathrm{km/s}}\right) & \left(\frac{m_b}{\Msun}\right)^{1/2}\left(\frac{m_\mathrm{per}}{\Msun}\right)^{-1} \left(\frac{a}{10^4 \mathrm{AU}}\right)^{1/2}\left(\frac{\log \Lambda}{5}\right)^{-1}.
\end{align}
This estimate suggests that $a_\mathrm{crit} \sim 10^4$AU is roughly the transition point between Galactic tide-dominated ($a > a_\mathrm{crit}$) and scattering-dominated ($a < a_\mathrm{crit}$) eccentricity dynamics. For $a_\mathrm{crit} \sim 10^4$AU --- which is a regime of key observational interest ---  it is not really legitimate to separate tidal effects from scattering. Nevertheless, by understanding the two effects individually we lay the ground for future work that will combine the two.

We have also assumed throughout this work that the initial distribution of wide binary orientations is isotropic. This assumption is very natural given that the scale height of the galactic disk is $\sim 200$\,pc while the widest binaries we consider here have semimajor axes less than $1$\,pc.   Equivalently, if one considers wide binaries formed through random pairing of stars in the field, then the typical velocity dispersions in any direction in the Galaxy are $\gtrsim 30$ km/s. Since the binary orbital motion is at a speed of $\sim 1$ km/s, the fact that the Galaxy has a ``velocity ellipsoid'' rather than a ``velocity sphere'' is unlikely to be of importance.

Another limitation of our model is that we have assumed a time-independent Galactic tidal potential, whereas in fact the Galaxy is evolving secularly, as are the orbits of the stars which comprise it (e.g. \citealt{mackereth2019dynamical}). However, in the present context this is unlikely to make much difference. Mathematically, we know that the Hamiltonian \eqref{eq:def_HGamma} with $\Gamma = 1/3$ is accurate as long as the disk is thin and the binary does not undergo vertical excursions that are larger than the disk scale height. Since orbits tend to migrate to larger vertical actions over time \citep{mackereth2019dynamical}, the binaries we measure today to be part of the thin disk have probably always satisfied this criterion. Any changes in the local density $\rho_0$ will therefore change the secular timescale \eqref{eq:t_sec_numerical} somewhat, but not fundamentally alter the characteristic dynamics.

Further, we have considered only a single species of binaries which were all born simultaneously at $t=0$. In reality, wide binaries have a range of ages, and so it is perhaps more realistic to take a continuous birth history over the previous $\sim 10$ Gyr. However, using a more realistic birth history would not change any of our key conclusions, as one can think of the full set of binaries as a superposition of families born at different times. The key findings of this paper (for instance, the fact that that tides will not turn a subthermal DF into a superthermal DF) will apply to each individual family, and thus to the whole population (since e.g. a superposition of subthermal DFs will also be subthermal). Nevertheless, one should keep in mind the fact that different families may have undergone different amounts of phase-mixing. Then, for example, older families will likely have an eccentricity DF that is closer to thermal than that of younger families.

Combined with the fact that the mixing timescale $t_0\propto a^{-3/2}$ is shorter for wider binaries, this effect may explain the observation that binaries with separations $10^3\,$AU are more superthermal than those with separations $10^4\,$AU \citep{Hwang21}. However, scattering by passing stars is such an important effect in the evolution of the widest binaries that at this stage, a detailed discussion of the implications of our results for these observational puzzles is premature. We therefore defer such a discussion to Paper II. Moreover, the results of this paper shed very little light on the other observational puzzle mentioned in the introduction, namely the unusually high eccentricities of twin wide binaries \citep{hwang2022wide}. Quadrupolar galactic tides are independent of the binary mass ratio, so we cannot explain why binaries with mass ratios close to unity should evolve differently from the others.

We also ignored the possibility that our wide binaries are in fact components of hierarchical triples. This was primarily because we were motivated by the work of \cite{Hwang21}, who isolated only those binaries without tertiary companions in their sample. In reality, a significant fraction of wide binaries likely \textit{are} part of triples \citep{Hartman2020}, and this adds another layer of complexity to the dynamics, since a wide triple interacting with Galactic tides comprises an effective hierarchical quadruple system \citep{Grishin2022}.

Finally, we emphasise that the results of \S\S\ref{sec:dynamics}-\ref{sec:coarse_graining} apply to arbitrary ensembles of Keplerian systems as long as they are perturbed by a common tidal Hamiltonian (so that the evolution of the ensemble's DF satisfies the Vlasov equation \eqref{sec:ensemble_evolution}), and this tidal Hamiltonian is sufficiently weak (so as to not change the orbits' semimajor axes, see \S\ref{sec:dynamics}). The results of \S\ref{sec:proofs} also apply to arbitrary Keplerian ensembles, with the additional constraint that the initial population of orbits must be oriented isotropically. As an illustration of a novel context in which our results might be of interest, consider the peculiar orbital distribution of young stars around Sag A* \citep{von2022young}. It has been suggested that this distribution might be produced by a secular perturbation due to an intermediate mass black hole (IMBH) \citep{zheng2020influence}. Our results would place constraints on the allowable mass and orbital parameters of such an IMBH for a given initial distribution of orbits (or vice versa), without needing to perform any numerical simulations. As another example, one could apply our formalism to the statistics of planetary eccentrities and inclinations as driven by Galactic and/or stellar cluster tides \citep{Perets2012, veras2013planetary, dupuy2022orbital}. We leave analysis of these and similar possibilities to future work.


\section{Summary}
\label{sec:summary}


In this paper, we have studied the evolution of an ensemble of wide binaries at fixed semimajor axis, evolving in the presence of a weak, smooth external tidal field (the Galactic tide). Our main results can be summarized as follows.

\begin{itemize}
    \item We extended the key result of \cite{Hamilton22}, namely that Galactic tides cannot on their own produce the observed superthermal eccentricity distribution of wide binaries, to arbitrary tidal Hamiltonians and arbitrary times.
    \item We used a specific model of the Galactic tide to probe numerically the evolution of various initial wide binary DFs. We found that a wide array of eccentricity DFs could be parameterized by a single ``effective power law index'' $\alphaeff$, which measures the total deficit or surplus of low-$e$ binaries compared to the thermal distribution.
    \item For the great majority of initial DFs, Galactic tides produce a final steady state eccentricity DF of power law form, on a timescale $ \sim 4\,\mathrm{Gyr}\,(a/10^4\mathrm{AU})^{-3/2}$.
    \item We discovered a close relation between the initial (effective) index $\alpha_\mathrm{i}$ and final index $\alpha_\mathrm{i}$ of the final eccentricity DF, namely $\alpha_\mathrm{f} \approx (1 + \alpha_\mathrm{i})/2$.
\end{itemize}

In a companion paper (Paper II), we will investigate the effect of stellar scattering on the combined semimajor axis and eccentricity DF of wide binaries. Incorporating insights from both papers will allow us to place constraints on the likely formation channels of wide stellar binaries in the Milky Way.


\section*{Acknowledgements}


We thank Yuri Levin for his suggestion to apply entropy arguments in this context, Robert Ewart and Michael Nastac for highlighting the utility of the Casimir invariant $C_2$, and George Wong, Jacob Nibauer, Nishant Mishra, Hsiang-Chih Hwang and Nadia Zakamska for further helpful discussions. We also thank the anonymous referee for valuable input. S.M. acknowledges support from the National Science Foundation Graduate Research Fellowship under Grant No. DGE-2039656. This work was supported by a grant from the Simons Foundation (816048, CH).


\section*{Data Availability}


The numerical simulation results used in this work will be shared on request to the corresponding author.




\appendix


\section{Estimating $C_2$ in numerical simulations}
\label{sec:C2_estimates}


Given a snapshot of phase space, we may estimate (using the fact that the phase-space average of a quantity $A$ is defined by $\langle A \rangle \equiv \int \md\bw Af$)
\begin{align}
    \label{eq:C2_estimate}
    {C}_2(t) \approx \frac{1}{N}\sum_{i=1}^{N} \widehat{f}(\bw_i, t),
\end{align}
where the index $i$ runs over all $N=10^5$ binaries. Following \cite{Beraldo_e_Silva_entropy}, we estimate the DF at the phase space coordinates of each binary using a kernel density estimate $\widehat{f}$, in this case with a 4D top-hat kernel. Note that this process does not involve any coarse-graining: we estimate the distribution directly from each binary's coordinates without averaging over any region of phase space. We confirm that the resulting estimated value ${C}_2$ is constant in time (to within finite-$N$ noise), as is expected for the evolution of a fine-grained distribution (equation \eqref{eq:C2_conserved}).

Next, we estimate the value of the \textit{averaged} quadratic Casimir (equation \eqref{eq:Casimir_Split_2}, where the average consists of an integral over $\omega$, $\Omega$ and $j_z$, as in equation \eqref{eq:f_avg_C2}):
\begin{align}
    \label{eq:C2bar_estimate}
    {\overline{C}}_2(t) \approx \frac{1}{N}\sum_{i=1}^{N} \frac{1}{8\pi^2j_i}\widehat{F}(j_i, t)
\end{align}
where $\widehat{F}$ is a kernel density estimate to the dimensionless angular momentum distribution $F$ using a 1D top-hat kernel. This quantity is what is shown in Figure \ref{fig:C2_vs_t}.

\bsp
\label{lastpage}
\end{document}